# Darwinian Genetic Drift

Alexey A. Shadrin[1,2], Dmitri V. Parkhomchuk[3, 4*]

[1]NORMENT, KG Jebsen Centre for Psychosis Research, Institute of Clinical Medicine, University of Oslo, Oslo, Norway.
[2]Division of Mental Health and Addiction, Oslo University Hospital, Oslo, Norway.
[3]Computational Biology of Infection Research, Helmholtz Center for Infection Research, 38124 Braunschweig, Germany.
[4]German Center for Infection Research (DZIF), partner site Hannover-Braunschweig, 38124 Braunschweig, Germany.
*Correspondence to: pdmitri@hotmail.com

**Abstract**

Genetic drift is stochastic fluctuations of alleles frequencies in a population due to sampling effects. We consider a model of drift in an equilibrium population, with high mutation rates: few functional mutations per generation. Such mutation rates are common in multicellular organisms including humans, however they are not explicitly considered in most population genetics models.
Under these assumptions the drift shows properties distinct from the classical drift models, which ignore realistic mutation rates: i) All (non-lethal) variants of a site have a characteristic average frequencies, which are independent of population size, however the magnitude of fluctuations around these frequencies depends on population size. ii) There is no "mutational meltdown" due to "low efficiency of selection" for small population size. Population average fitness does not depend on population size. iii) Drift (and molecular clock) can be represented as wandering by compensatory mutations, postulate of neutral mutations is not necessary for explaining the high rate of mutation accumulation.
Our results, which adjust the meaning of the neutral theory from the individual neutrality of the majority of mutations, to the collective neutrality of compensatory mutations, are applicable to investigations in phylogeny and coalescent and for GWAS design and analysis.

**Introduction**

Before the discovery of DNA and invention of sequencing, researchers assumed that mutations are rare and accumulate mostly in one-by-one fashion. This assumption came from the observations that visible phenotypic mutants are rare and mutants, which have increased fitness are even more rare. Now we realize that most DNA mutations do not produce readily observable mutant phenotype, but have more subtle, but functional consequences (*e.g.* variability of metabolism and so on). However, the mentioned observational illusion produced a common view that selection can evaluate mutations individually, one-by-one. This assumption shaped the start of population genetics models, and heavily influenced subsequent developments up to present time. Early experiments with bacteria also indicated rather low per-generation mutation rates. In this case a negative mutation is promptly wiped out, while a positive mutation go through a "fixation" process: an increase of frequency until the whole population carries it. Haldane pointed out that the fixation rate has inherent limits (Haldane, 1957): *e.g.* to attain a fixation in a population with one million individuals, a mutation has to multiply accordingly through many generations. And this process must happen to every positive mutation. Although some number of positive mutations can be multiplied in parallel in a population, and eventually put together by genetic recombination, the process has severe throughput limits. After the DNA discovery and genes sequencing, it was found that there are too many accumulated mutations between species. Normally, one would try to develop a theory which handles variants with arbitrary degrees of functionality, and then, if necessary, one can put functionality to zero to observe the behavior of "neutral" variants. However due to the views that negative mutations are promptly rejected, while positive are sent through the fixation process with the limited rate, the proposed solution was ubiquitous "neutrality". Such a solution also rescued the prevailing one-by-one fixations paradigm. As stated in the corresponding founding papers: "Calculating the rate of evolution in terms of nucleotide substitutions seems to give a value so high that many of the mutations involved must be neutral ones" (Kimura, 1968) or: "Most evolutionary change in proteins may be due to neutral mutations and genetic drift" (King and Jukes, 1969). The logic of these statements holds under the assumption that selection can evaluate mutations one-by-one. If we consider that there are few functional mutations per generation, the assumption is completely violated and this reasoning becomes inapplicable and cannot be rescued. It seems, the wide popularization of "neutrality" was an emotional reaction: "neutrality" ("blind" force) was viewed as an opposition to Darwinian selection ("directional" force). Hence the "non-Darwinian" label was coined for such phenomena. Interestingly, emotions were quite apparent on the path of these models (Gillespie, 1984). We suggest that more reasonable opposition to the selection force is mutagenesis (Fig. 1).

Later, it was realized that per-genome per-generation mutation rate in humans is around 70 mutations, with significant variation, caused by parents' age and other factors (Keightley, 2012). Notably, human (per nucleotide) mutation rate is lower than that of primates and other mammals. The question of a "functional fraction" of a human (or of any large) genome is a very difficult one for current theories, because the relationship between functionality and sequence conservation is poorly understood. However, if we take, conservatively, that about 10% of a genome show conservation (Ponting and Hardison, 2011), then at least this fraction of random mutations is functional, and we have at least seven functional mutations per generation. Instead of trying to patch previous theories to include high mutation rate, we prefer to build a new one without any "neutrality" assumptions, pure neutrality is clearly an idealization. Of course, within this model, we can explore "neutrality", if interested, just by putting functionality to zero.

The presented model is minimalistic in a sense that it cannot be made more simple. On the other hand it is easy to make it more complex through the inclusion of additional factors, such as epistasis, epigenetics, ploidy and so on. We restrain ourselves from in-depth comparisons of this model to many other models, as they have different applicability domains, due to the different starting assumptions. As soon as the high rate of functional mutations is assumed, the "meltdown" of a "perfect" genome is inevitable (Fig. 1), with the consequences we discuss below. We are not aware of any model, which acknowledges this fundamental "meltdown" with similar formalism. Practically, the only approach with appreciable mutation rates is the quasi-species model (Eigen, 1971), which has important differences with our treatment.

Regrettably, nearly a hundred years old considerations about population size (*e.g.* its influence on fitness), which were based on the presumed prevalence of one-by-one mutations accumulation mode, found their ways into population genetics textbooks, under the pretense of being universal principles, which are applicable to any population. For example: "From a modest beginning, when Sewall Wright dealt with the process of genetic drift in a population with two sexes, the concept of effective population size has been extended to the status of a unifying principle that encompasses the action of drift in almost any imaginable evolutionary scenario." (Charlesworth, 2009). On the other hand consider Wright's views on mutagenesis: "The observed properties of gene mutation—fortuitous in origin, infrequent in occurrence and deleterious when not negligible in effect—seem about as unfavorable as possible for an evolutionary process." (Sewall Wright, 1932). We cannot resist to quote Thomas Henry Huxley: "Mathematics may be compared to a mill of exquisite workmanship, which grinds you stuff of any degree of fineness; but, nevertheless, what you get out depends upon what you put in;...". What are the chances that formalisms (and far-reaching interpretations) derived from these early views, will luckily coincide with a formalism required to describe quite high mutation rates? Scholars of population genetics should be aware, that classical models assume negligible mutation rates, while for realistic (per-generation) mutation rates (*e.g.* that of mammals) the formalism should be very different. Roughly speaking, in the later case, the mutagenesis is a powerful force, which prevents selection from attaining a "perfect" genome. However, the frequency of beneficial mutations is significant, precisely due to the "degraded" state of a genome. The drift (or population size) has no influence on population fitness (inbreeding effects are unrelated to drift, and do not happen in an equilibrium haploid population we consider here).

**Methods and Results**

We assume that the environment is constant and we know a "perfect" genome for this environment. A perfect genome is not observed in realistic conditions, it is used here as a starting point for modeling purposes. To aid a perception we assume that "A" is the best nucleotide in all positions (Fig. 1). We assume that each nucleotide has selective weight ("A" having the highest weight), which defines its contribution to individual fitness. At first, we consider all positions having the same weights set. Formula for individual fitness can be a sum or multiplication of sites' weights, we assume no epistasis, selection simply tries to attain the best variant ("A") in every position. We also do not consider lethal sites, which are invariable. For specificity, we assume that each newborn individual receives ten mutations in average. If we let a population of "perfect" genomes to evolve, then at the first generation each individual will get ten deleterious mutations in average. In a stable population, the average number of (genetically) surviving children is one per parent, the rest constitute "genetic deaths". Lets assume ten children for a parent (20 for a couple in a recombining sexual populations). All of them receive ten *de-novo* mutations in average. However at least one child per parent, in average, must have the fitness of the original population, for the equilibrium condition. For the perfect starting genomes, there is no way selection can remove these ten deleterious mutations, hence in the subsequent generations deleterious variants will accumulate further, and fitness will drop. The initial population of perfect genomes will undergo "mutational meltdown". However, the positive side of this meltdown is that at some point significant amount of suboptimal sites will accumulate in genomes, and at this point some of these ten *de-novo* mutations will be positive (from "G" to "A", for example). The equilibrium condition can be met at some point, because when a large fraction of a genome is suboptimal (far from the "perfect" genome), a significant fraction of random mutations is positive. Naturally, the equilibrium level depends on reproductive capacity and mutation rate. Apparently, such maintenance does not depend on population size unless it is very small. The process, which maintains fitness from dropping, due to random mutagenesis, is the selection from a litter, of at least one child per parent (in average), having the fitness of the main population. A litter should be understood in a general way, it can be a pooled litter, *e.g.* from neighboring parents. An individual is evaluated by selection only locally, in competition with neighbors, selection can not evaluate him against the whole population, unless it is very small. This process maintains the average population fitness, compensating the per-generation average fitness drops. Naturally, waves of selective sweeps can overlap with this process, but they are rare in comparison with per-generation fitness maintenance.

There are some estimates on the number of deleterious mutations per generation in human genome: for example an estimate of 2.2 deleterious mutations (Keightley, 2012). However, to make this number meaningful, one must also provide an estimate for the number of positive mutations. Again, we can see the legacy of foundations of population genetics: positive mutations rate is "negligible". However the number of positive mutations cannot be irrelevant for this estimate, and as we try to show here it is not negligible: what if we have 2.2 negative and 3.3 positive mutations, in average? Such situation seems "unthinkable" in the current paradigm, however we can easily construct such a genome, which has many suboptimal sites, where random mutations are mostly positive. Conventional wisdom dictates that positive mutations are (unspecifically) "rare". For example: "mutations having negative or negligible effects on fitness are more common and thus easily studied, while those having positive effects on fitness are far rarer and thus studied only with difficulty." (Orr, 2010). In our model, the numbers of (*de-novo*) positive and negative mutations are comparable, however their specific values are not important for the points we discuss here.

Models of drift (Charlesworth, 2009), which claim that small population size is detrimental and leads to "mutational meltdown" (which is different from our static "meltdown") do not consider realistic mutation rates, usually they consider the drift without mutagenesis at all, hence the applicability domain is unrealistic. If we let an initially variable population to drift without mutagenesis, it will eventually become monoclonal, how real is that? Why do we want to model this unrealistic effect, instead of making more realistic model with persisting variability? Naturally, in such models, without mutagenesis, the population size is important: for example it determines how fast the initial variability is lost, it influences the probability of fixation of bad or good variants. Without mutagenesis the fixation is a final event for any variant. Such formalism shows that a bad variant is more likely to be fixated in smaller populations. This behavior simply reflects the starting assumptions that the mutagenesis can be "neglected" or considered somehow "separately" from drift effects. While mathematically correct for their assumptions, these models do not address the questions: How numerous *de-novo* functional hitchhikers, appearing at every generation, might influence the fixation process? What is the rate of "un-fixation" (by *de-novo* mutations) of the site in question? An attempt to answer these questions would inevitably result in a model analogous to the presented here. A fixation is not a single final event in the life of a site. Non-lethal variants are going through continuous rounds of fixations and "un-fixations" in realistic conditions. When the population size is sufficiently large, and hence the frequency of "un-fixations" is high, a (non-lethal) variant is never fixated, but is characterized by its average frequency in a population. In a small population this average frequency can be revealed with sufficiently long time average ("ergodic" property). In this model, potential variability and fitness does not depend on population size. Inbreeding effects, which might play some role for small populations, are quite different phenomena, they are not considered here.

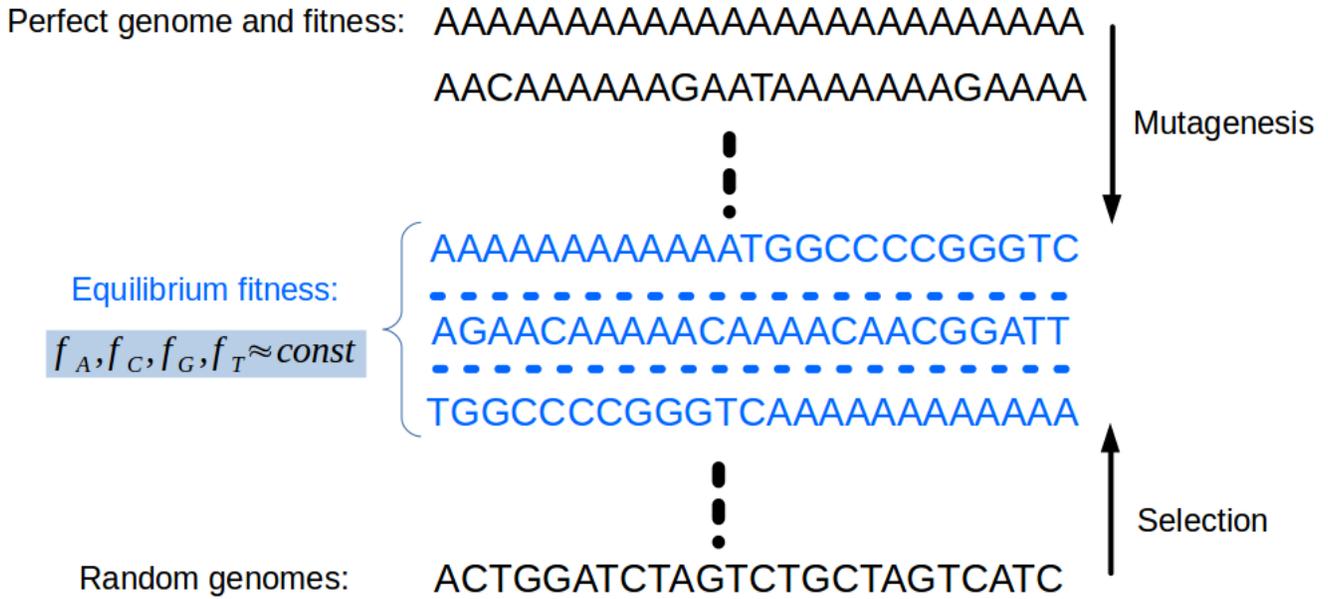

**Fig. 1.** Population of perfect genomes "melts down" to an equilibrium population.

Fig. 1 illustrates the equilibrium population (blue). What is the possible number of distinct genomes in this population? As an approximation, let's first consider strictly constant fitness and hence constant nucleotide frequencies in each genome. In this case the fitness is invariant with respect to any sequence permutations. Assuming $L$ is a constant length of genome in the population, the total number of permutations in a genome is $L!$, to determine the number of distinct possible genomes, in this population, we have to divide this number by the number of redundant permutations (between two "A"s, etc.). Hence the total number of possible genomes is described by multinomial formula:

$$\frac{L!}{A!C!G!T!} \quad (1)$$

where A, C, G and T are the numbers of corresponding nucleotides in a genome at average fitness.
This number (1) is apparently very large and is not convenient for handling, *e.g.* its connection with expected variants' frequencies in a population is obscure. This number can be compared to the number of available micro-states in statistical mechanics. Lets follow the recipe from statistical mechanics and calculate a logarithm of this number normalized by a genome length. That will give some per-site measure of potential population variability. Put more precisely, it quantifies the variability, which is available for a population to explore, while the actual observable variability can be smaller, as it is represented by a limited population.
Applying Stirling's approximation for factorials:

$$\frac{1}{L}\log_2 \frac{L!}{A!C!G!T!} \approx \frac{1}{2L}\left(\log_2(2\pi L) - \sum_{n \in A,C,G,T} \log_2(2\pi n)\right) - \sum_{n \in A,C,G,T} f_n \log_2(f_n), where: f_n = \frac{n}{L}$$

$$= -\sum_{n \in A,C,G,T} f_n \log_2(f_n) + O\left(\frac{\log_2 L}{L}\right)$$

(2)

$$H(f) = -\sum_{n \in A,C,G,T} f_n \log_2(f_n) \quad (3)$$

$$GI(f) = 2 - H(f) \quad (4)$$

Hence, the variability per site, is conveniently described by the entropy formula (3) applied to the expected variants' frequencies. In a GC-balanced genome, we can also define genetic information (*GI*) as the value reciprocal to entropy (4), so highly conserved (invariable) site has the maximum of 2 bits of information, while a non-conserved site has 0 bits. In permutations approximation, all sites are described by the same expected frequencies, in this formula. However, in reality, sites can have different "degree of conservation": their sets of expected frequencies can be different. Typical set, one of the fundamental concepts of Shannon's Information Theory (IT), allows to generalize our simplified permutations approach, with flat conservation, to an arbitrary conservation profile. Then, each site in a genome, can have different expected frequencies and variability, depending on its conservation. Typical set represents the set of possible genomes for corresponding conservation profile.

The permutations approximation of a typical set has mostly educational value: technically we could jump straight to the typical set description of this model. However, without prior knowledge of mathematics of IT, it might be difficult to grasp immediately what is a "typical set" and why it is introduced here, as if the appeal to IT formalism is artificial and "obfuscates" the model. The appearance of IT mathematical tools is a mathematical "coincidence" here; similarly, the Law of Large Numbers appears "independently" in different unrelated models. Genetic information can not be easily compared to ordinary information, *e.g.* that of a memory stick, though one can compare *GIs* of different species or genes, for example. The permutations set, which is more comprehensible, serves as an intermediate step, an approximation to the typical set. Correspondingly, a permutation represents two compensatory mutations (one positive and one negative), while wandering (drifting) in a typical set is performed through arbitrary combinations of compensatory mutations, *e.g.* one "strong" mutation can be compensated by few "weak" mutations. When one is not sure about the meaning and properties of a "typical set", one could use "permutations set" as an approximation.

**Discussion and conclusion**

Formalism shown in Fig. 1 can be applied not only to the whole genome, but for individual gene sequences (which take a corresponding share of genetic deaths for maintenance), or instead of nucleotides one can consider variants of larger genetic units (*e.g.* genes or codons). As can bee seen (Fig. 1), the principal core of the model is parameter free: if the mutation rate is high enough, so that the equilibrium "departs" from the "perfect" genome, the combinatorial effects and other properties we discuss here hold, regardless of the specific position of the equilibrium. For example, adding epistasis, (di)ploidy, recessiveness/dominance, genetic code into considerations will add some interesting effects, for example, masking deleterious recessive variability can be advantageous for fitness. However the general "combinatorial" properties and "compensatory mutations" description will remain, although in more complex form. Therefore, we assume that these properties are inevitable in this applicability domain (high functional mutation rates). Any other population genetic model in this domain must have such properties. There are qualitative differences between high and low mutation rates models: when the mutation rate is not sufficient for the "melting" of a "perfect" genome (Fig. 1), our description is non-applicable. On the other hand, many classical models with insufficient mutation rates are merely

"scratching the surface", as if genomes are fluctuating near the "perfect" genome (Fig. 1), without going "deep" into variability. These two types of models are, thus, incompatible. Some classical concepts are not directly transferable into high mutation rate domain. For example, the concept of "selection coefficient" (which quantifies the reproductive success of a variant) is not applicable in equilibrium population. Good and bad variants in a large population maintain their constant average frequencies (hence, "paradoxically", their *de facto* selection coefficients are zero), while in a small population frequencies are drifting around these average values. Since the drifting variants are not neutral, and their average frequencies are deviating from 0.25 (expected for neutral variants in a GC-balanced genome) we can call such a drift "Darwinian": Darwinian selection is offsetting the drift (its average frequencies) of all variants. In the case of environment changes, these average frequencies might change, however in that case the "selection coefficients" are changing in time (towards zero), during the return to the equilibrium. In our case a variant is defined by its "weight", which defines how it contributes to individual fitness, it is constant in time (in constant environment).

Naturally, any model has its purpose and value in explaining and predicting empirical phenomena. Therefore we briefly mention some of phenomena, which seem to be easier explainable in this framework. We consider explanations for molecular clock, Drake's rule, advantage of genetic recombination and the capacity to maintain functional sequences with nearly neutral substitutions rates. We remind, that a typical set determines population (potential) variability and fitness, population size only defines the "volume" occupied by population in this typical set. Hence we can explore how various operations (*e.g.* change of a genome size, mutation rate, recombination) affects the typical set, and can make inferences about fitness of a population of any size, which "lives" in this set.

Usually, molecular clock is explained (away) with the neutral theory. The provided model suggests that molecular clock is an accumulation of compensatory functional mutations (collective neutrality rather than individual neutrality). Obviously, our approach is more suitable to explain why the clocks are ticking with different rates in strongly and weakly conserved genes, without creating any awkward concepts of differential "neutrality density", for example. It also indicates why molecular clock does not depend on population size, despite operating with functional variants.

Drake's rule is a simple consequence of maintaining constant functionality, sites' conservation and variability during the changes of functional genome size. The equilibrium region in Fig. 1 (blue) should not change, during the change of genome size, to preserve genes' functionality at the same level. As we discussed, the limiting step of selection is to maintain at least one successor per parent at the average fitness. Assuming that the reproductive capacity is more or less conserved for similar species (and taking into considerations that the rule holds on the log-log scale), when a functional genome length is increasing, mutation rate must decrease to keep the number of mutations per genome per generation constant, otherwise the variability will increase and sequence functionality will degrade (Shadrin and Parkhomchuk, 2014).

Fig. 1 shows an example of a typical set (blue), in a "sorted" order. All sequences there have the same fitness and can coexist in an asexual population. In a sense, genetic deaths are "wasted" to maintain coexisting upper and lower genomes. It is apparent that, if we add recombination operation in this set, genetic deaths can be exploited more efficiently, because, for example, recombination of the upper and lower genomes might produce a perfect genome (and a very bad genome, discarded by selection), increasing the frequency of "A"s. In average, the effect of random recombinations is smaller than this extreme example, however it decreases the size of the typical set and increases the fitness: the frequencies of "A"s. It is easy to develop more formal approaches (to be published elsewhere) or to perform numerical experiments elucidating this phenomenon (Parkhomchuk *et al.*, 2016). Interestingly, at the typical set level, recombination decreases variability (the set is smaller), while on the individual (or limited population) level, one can perceive an increase of variability (*i.e.* a genetic distance from parents to children, as compared to parthenogenesis). When two equally fit parents, with different genomes, recombine, the fitness of progeny is distributed equally above and below parents fitness. Then selection can pick the progeny with the increased fitness and discard the opposite. This explanation looks trivial, and calls for the question of why it was not suggested before, while dozens of other hypotheses were put forward, for this important evolutionary problem. However, as we discussed in the Introduction, this trivial explanation requires the acceptance of the general features of our model: 1) The mutation rate is not "negligible", in fact, its "force" is equal to the opposing selection "force" (Fig. 1). Thus, exploring selection (drift, recombination and so on) effects without explicitly including the balance with mutagenesis is critically incomplete; 2) A large fraction of a genome is suboptimal, far from a "perfect" genome, inevitably contaminated with deleterious variants. Genes are not near their "optimal performance", in general. The frequency of positive mutations is high. There is some experimental evidence for this view (Hall and Joseph, 2010).

From the above considerations, it is apparent that for the more variable (larger) typical set the effect of recombination is more pronounced (Fig. 1). For monoclonal population (typical set of size one) the recombination is useless, while for the maximum variability, recombination advantage is also maximized. Near the maximum of variability, sequences have near neutral substitution rate and can be only weakly functional with genetic information (*GI*) approaching zero, in asexual populations. However, the presence of recombination boosts the capacity to maintain relatively high *GI* for such sequences. Individual variants have weak effects in this case, and arrive in a genome in "batches" (not one-by-one), so selection is acting on them not individually (aka purifying selection) but on their combinations, continuously created by recombination. That resembles Muller's ratchet idea (Felsenstein, 1974), however we complement it with proper inclusion of positive (*de-novo*) variants. Without this inclusion, one would have to postulate a very specific form of epistasis for negative mutations, to get this explanation of recombination advantage to work. As individual negative mutations with weak effects are not rejected immediately, the substitution rate appears to be nearly neutral, while *GI* can be relatively high (~1 bit). For weakly conserved sequences the frequency of beneficial mutations is about 50%, that could be the substitution rate if all deleterious mutations were rejected immediately. However, since a significant fraction of them is allowed to stay in a population for a prolonged period, the rate can appear to be nearly neutral. This prediction addresses the "junk DNA" paradox. Weakly conserved sequences (constituting ~90% of human genome) might be not an evolutionary "waste" or "junk". To the contrary, they seem to be a more efficient way of storing genetic information, given high per-genome mutation rates. More elaborate demonstration of this effect deserves separate investigation, however an interested reader can easily investigate it with simple numerical experiments or more formal approaches.